\documentclass[lettersize,journal]{IEEEtran}
\usepackage{amsmath,amsfonts}
\usepackage{algorithmic}
\usepackage{algorithm}
\usepackage{array}
\usepackage[caption=false,font=normalsize,labelfont=sf,textfont=sf]{subfig}
\usepackage{textcomp}
\usepackage{stfloats}
\usepackage{url}
\usepackage{xurl}
\usepackage{seqsplit}
\usepackage{verbatim}
\usepackage{graphicx}
\usepackage[T1]{fontenc}
\usepackage[utf8]{inputenc}
\usepackage{hyperref}
\usepackage{booktabs}
\usepackage{listings}
\usepackage{xcolor}
\usepackage{balance}
\usepackage{todonotes}
\usepackage{placeins}

\lstdefinelanguage{yaml}{
	keywords={true, false, null, yes, no},
	keywordstyle=\color{blue},
	basicstyle=\ttfamily\scriptsize,
	comment=[l]{\#},
	commentstyle=\color{gray},
	stringstyle=\color{black},
	morestring=[b]',
	morestring=[b]"
}

\title{B.O.D.Y. — Beyond-Overlay Deterministic topologY: A Layer-2 Declarative Ground Truth for AIOps Pipelines under Fragmented Administrative Boundaries}

\author{Matheus~Marques,  Carlos Alberto Malcher, Cledson~de~Sousa%
	\thanks{C.~de~Sousa is with GTECCOM and the MídiaCom Laboratory, Universidade
		Federal Fluminense (UFF), Niterói, RJ, Brazil.
		E-mail: cledson@id.uff.br}%
	\thanks{This work was conducted within the scope of
		\textit{Projeto Monitora} at UFF. 			The monitoring infrastructure --- 1,000~IP cameras and
		30~Linux servers across five \textit{campi} --- was operated by GTECCOM/UFF, which provided the operational dataset. The shared GPU inference infrastructure (NVIDIA H200~NVL)
		was provided by MídiaCom/UFF.}}
\begin{document}
	
	\maketitle
	

\begin{abstract}
	Modern AIOps environments operating within multi-campus institutional 	infrastructures suffer acutely from topological drift and 	\textit{black-box} unmanaged physical network segments. Classical
	layer-2 discovery pipelines rely on uniform administrative cooperation 	and ubiquitous SNMP polling, which routinely fail in heterogeneous, 	multi-vendor, and multi-tenant overlay infrastructures. This paper
	introduces \textsc{B.O.D.Y.} (Beyond-Overlay Deterministic topologY), a 	deterministic structural grounding layer for AIOps ecosystems operating 	under fragmented administrative boundaries. B.O.D.Y. bypasses the
	NP-hard incomplete Address Forwarding Table (AFT) resolution dilemma 	by formalizing a multi-modal data fusion pipeline that orchestrates 	ephemeral MAC address forwarding tables collected via non-privileged
	terminal sessions, passive OUI fingerprinting, PoE telemetry, and 	declarative state storage. The resulting topology graph reconstructs 	unmanaged physical segments and maps logical asset semantics without
	administrative network privileges, providing downstream reasoning 	systems with an immutable, auditable source of physical ground truth. 	Evaluation across five campuses of the Universidade Federal
	Fluminense, resolving 530 of 541 registered edge devices, demonstrates 	that deterministic topological grounding eliminates a critical failure 	mode in probabilistic AIOps reasoning: confident causal inference
	decoupled from physical reality.
\end{abstract}


\section{Introduction}

Accurate physical topology is the essential precondition for automated root-cause analysis in AIOps environments~\cite{notaro2021survey,rca2024automatic}. Without an auditable map of which device is connected to which port, on which floor, and under which power delivery constraint, probabilistic inference engines operating on metric streams and log pipelines produce what We term this failure mode \textit{topological hallucination}: the
production of confident causal inferences by an AIOps reasoning engine that are decoupled from the physical network state, in direct analogy to factuality hallucinations in large language models~\cite{hallucination_survey_2025}.

The canonical layer-2 topology discovery pipeline, thoroughly surveyed
by Ahmat~\cite{ahmat_topology_survey} and extended by Farkas et~al.~\cite{farkas2008topology},
relies on a critical and brittle assumption: comprehensive SNMP access across
all switching fabrics to parse Address Forwarding Tables~(AFTs). Incomplete AFT
resolution in heterogeneous single-subnets has been proven by Gobjuka and
Breitbart~\cite{gobjuka2010} to be an NP-hard problem, and the operational
conditions that produce incomplete AFTs are not edge cases: they are the default
condition in any overlay infrastructure operating under fragmented administrative
boundaries.

The obstacle is not purely technical. In many institutional environments,
the switches aggregating monitored traffic are managed by a separate networking
team whose cooperation cannot be assumed; PoE edge switches are routinely
provisioned without SNMP communities; and in multi-campus organizations where
each campus operates under a distinct administrative unit, even read-only SNMP
access to the full switching path is not guaranteed. Campus surveillance networks
exemplify this failure mode acutely: IP cameras do not speak SNMP, unmanaged
PoE switches are ubiquitous at the edge, and physical wiring documentation,
where it exists at all, is notoriously stale. The deployment environment
evaluated in this work — a federal university spanning five campuses across
distinct administrative boundaries — is representative of this class of scenario
rather than an exception to it. The same operational profile applies to any
overlay infrastructure operating on a dedicated VLAN with a semantically coherent
device fleet: wireless access point networks, industrial IoT deployments, VoIP
fabrics, and physical access control systems all encounter the same structural
constraint.

To overcome this mathematical and operational impasse, this paper introduces
\textsc{B.O.D.Y.}~(Beyond-Overlay Deterministic topologY), the deterministic
structural foundation of the \textsc{H.U.M.A.N.} framework. Instead of
treating topology discovery as a pure active network polling exercise,
\textsc{B.O.D.Y.} reconstructs layer-2 network graphs by fusing MAC address
forwarding tables collected via non-privileged CLI sessions with passive OUI
fingerprinting, PoE telemetry, and declarative state storage — bypassing the
incomplete-AFT problem without requiring SNMP access or cross-departmental
cooperation. The resulting topology graph is immutable, auditable, and resilient
to upstream infrastructure failures, establishing the physical ground truth on
which the \textsc{B.R.A.I.N.} cognitive inference layer~\cite{brain2026} operates.

The primary contributions of this work are:

\begin{itemize}
	\item A multi-modal topology inference pipeline that reconstructs layer-2
	network graphs without administrative switching fabric cooperation, fusing
	PoE telemetry, OUI fingerprinting, and MAC density analysis with declarative
	asset configurations.
	
	\item An idempotent, declarative filesystem hierarchy that persists topology
	state as an auditable source of truth independent of database or network
	availability~\cite{morris2020,hellerstein2010}.
	
	\item A structured Human-in-the-Loop protocol that exposes unresolved MAC
	addresses as named topology nodes, equipping field technicians with maximum
	context before physical intervention.
	
	\item Production validation across five campuses of the Universidade Federal
	Fluminense, resolving 530 of 541 registered assets with 97.9\% accuracy
	without administrative privileges on the institutional switching fabric.
\end{itemize}

The remainder of this paper is organized as follows.
Section~\ref{sec:related} surveys the topology discovery and asset
identification literature. Section~\ref{sec:architecture} details the
\textsc{B.O.D.Y.} architecture. Section~\ref{sec:results} presents the
production deployment results. Section~\ref{sec:limitations} discusses
limitations and Section~\ref{sec:conclusion} concludes.

\section{Related Work}
\label{sec:related}

The operational and theoretical landscape surrounding the \textsc{B.O.D.Y.} framework intersects three research domains: classical layer-2 topology discovery and its fundamental dependence on administrative cooperation;
asset identification in uncooperative environments and the limits of passive sensing modalities; and the emerging integration of Large Language Models into network management, where deterministic topological grounding emerges as an unaddressed prerequisite.

\subsection{Layer-2 Topology Discovery: The Classical Pipeline}

The canonical Ethernet topology inference pipeline relies on a deceptively simple premise: collect Address Forwarding Tables~(AFTs) from SNMP-enabled switches and reconstruct physical connectivity from the resulting reachability matrix. Lowekamp et~al.~\cite{lowekamp2001topology} formalize this pipeline and prove that topology discovery from incomplete AFTs is fundamentally dependent on SNMP access across all switching paths. Breitbart et~al.~\cite{Breitbart2004NetInventory} and Pandey et~al.~\cite{Pandey2011SNMPTopology} extended this foundation to heterogeneous multi-vendor deployments, while subsequent  efforts~\cite{Sun2006CRTTopology,Tao2009LinkLayerSNMP,Son2005MetroEthernet,Wang2016SNMPAlgo,Zhang2017SNMPOptimization} refined scalability and convergence within the same assumptions.Gobjuka and Breitbart~\cite{gobjuka2010} establish the theoretical ceiling: topology discovery from incomplete AFTs is NP-hard, and the incompleteness that triggers this complexity is not an edge case but the default condition in any overlay infrastructure operating under fragmented administrative boundaries. 

The collective achievement of this body of work is substantial within  operational assumptions. The critical proviso is uniform administrative cooperation across the entire switching path, an assumption the literature treats as an engineering constraint rather than a fundamental limitation~\cite{ahmat_topology_survey}. B.O.D.Y. departs from this paradigm  by replacing continuous active SNMP polling with a non-invasive multi-modal fusion pipeline that operates without administrative privileges, bypasses AFT incompleteness through PoE telemetry and OUI fingerprinting, and persists topology state as a declarative filesystem hierarchy rather than a relational database.

\subsection{Asset Identity and Topological Integrity in Uncooperative Environments}

Passive MAC address correlation without structural context reduces to probabilistic proximity heuristics within the MAC address   space~\cite{niedermaier2019}. Deterministic alternatives exist but require   protocol-level modifications to the switching fabric~\cite{hadzic2001},   precisely the form of administrative cooperation unavailable in fragmented   overlay environments. Both approaches expose the same gap: maintaining   topological integrity over time in uncooperative environments where physical   assets are replaced or provisioned outside the monitored administrative domain.

Imamura et~al.~\cite{imamura2020} quantify this gap empirically: a weighted combination of six passive identification methods achieves 78.4\% overall accuracy, a 23.0\% gain over any individual method. The structural ceiling, however, is revealing. Switching hubs achieve 0\% identification accuracy across all six methods combined, because passive traffic-based approaches cannot distinguish infrastructure devices that generate no application-layer traffic of their own. B.O.D.Y. dissolves this failure mode not through a single sensing modality but through a deterministic classification cascade (Fig.~\ref{fig:bodypipeline}): LLDP neighbor identity discriminates uplinks from managed cascades; PoE delivery state separates powered endpoints from passive infrastructure; MAC density on a single port reconstructs unmanaged cascade segments invisible to any management protocol; positional hostname semantics partition the resolved endpoints into logical floor blocks; and OUI
prefix lookup within the constrained manufacturer subspace of the monitored infrastructure assigns device class deterministically. Unresolved assets are not discarded but exposed explicitly in the topology graph, preserving full observability of registry integrity.

\subsection{LLM-Based Network Management and the Topological Grounding Gap}

The integration of Large Language Models into network management has    accelerated rapidly, but the literature converges on a structural    prerequisite that remains unaddressed. Hong et~al.~\cite{hong2025llm_netmgmt}    survey the landscape and identify hallucination and domain adaptation as    persistent open limitations. Hamadanian et~al.~\cite{hamadanian2023holistic}    propose a modular incident management architecture and note explicitly that    LLMs can diagnose simple network problems but fail on complex ones,    identifying network topology understanding as a required capability that    current models do not reliably exhibit. Neither work specifies how that   topology is to be obtained in environments where classical discovery fails.

Donadel et~al.~\cite{donadel2024llms} make the dependency concrete: in a    controlled evaluation across six LLMs and three topologies of increasing    complexity, accuracy on topology-dependent tasks drops systematically with    network size, and models fabricate connections that do not exist in larger    topologies. The best-performing model achieves 79.3\% average accuracy only    because the topology is provided as structured input. The accuracy figures    therefore represent an upper bound on what is achievable when ground truth    is available, not a characterization of what happens when it is absent or    stale. B.O.D.Y. operates precisely in that uncharacterized regime.

GeNet~\cite{genet2025} instantiates the same dependency architecturally:    its first module converts a topology image into a structured textual    representation before any intent can be processed. Intent implementation    quality is largely insensitive to topology image quality, provided the    topology is present. The failure mode GeNet does not address is the absence    of a topology representation altogether, which is the default condition in    overlay infrastructures operating under fragmented administrative boundaries.   \textsc{B.O.D.Y.} produces the deterministic topology that both  assume but neither provides, closing the grounding gap at the physical
layer before any cognitive inference begins. Its output spans three complementary representations: a visual graph for operator-facing triage, a declarative filesystem hierarchy as an auditable source of truth, and a per-switch tree structure for downstream reasoning pipelines. The architecture that realizes these outputs is detailed in the  section \ref{sec:architecture}.

\section{The B.O.D.Y. Architecture}
\label{sec:architecture}

\begin{figure*}
	\centering
	\includegraphics[width=1\linewidth]{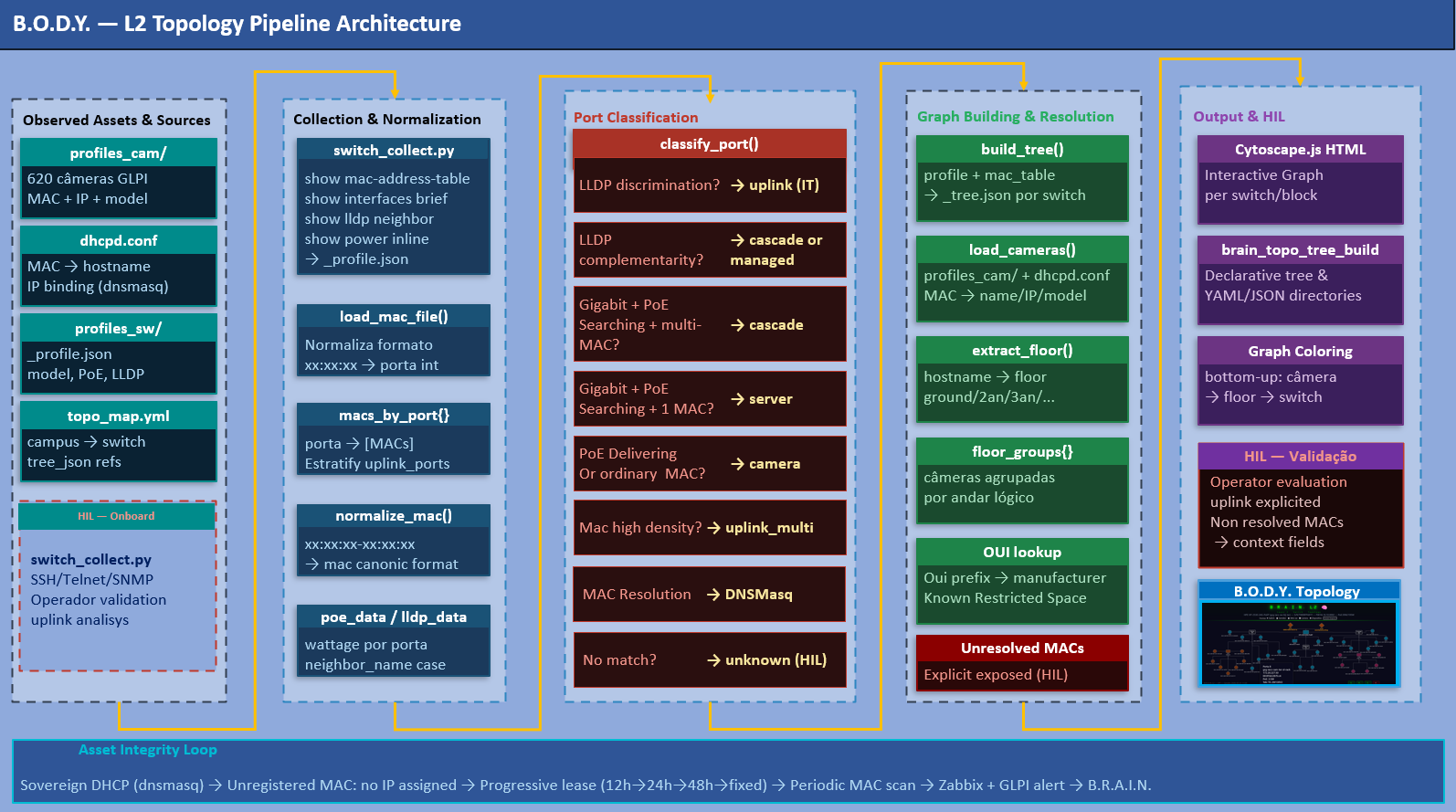}
	\caption{B.O.D.Y. L2 Topology Pipeline. From left to right: asset sources and HIL-assisted onboarding; non-privileged CLI collection and MAC normalization; deterministic port classification by LLDP case, PoE state, and MAC density with inline dnsmasq resolution; floor-grouped graph construction with explicit unresolved MAC exposure; and interactive topology output with bottom-up graph coloring. The Asset Integrity Loop enforces sovereign DHCP control and progressive lease stabilization throughout the asset lifecycle.}
	\label{fig:bodypipeline}
\end{figure*}

The preceding survey establishes three compounding constraints: classical AFT-based discovery fails without uniform SNMP cooperation~\cite{gobjuka2010}; passive identification methods cannot classify infrastructure devices that generate no application-layer traffic~\cite{imamura2020}; and LLM-based management frameworks presuppose a structured topology that neither classical nor passive approaches can reliably produce in fragmented overlay
environments~\cite{donadel2024llms,genet2025}. \textsc{B.O.D.Y.} addresses these constraints jointly by rejecting
continuous active probing in favor of a non-privileged, on-demand multi-modal fusion pipeline that processes heterogeneous telemetry at the single programmable edge distribution switch boundary. Fig.~\ref{fig:bodypipeline} illustrates the complete pipeline architecture, organized into five functional stages detailed in the subsections that follow: asset consolidation from canonical data sources, multi-vendor CLI collection and normalization, deterministic port classification, graph construction and resolution, and interactive topology output with structured HIL intervention. The Asset Integrity Loop, shown at the base of the figure, enforces sovereign DHCP control and progressive lease stabilization throughout the asset lifecycle independently of all pipeline stages.

\subsection{Observed Assets}
\label{sec:assets}

\textsc{B.O.D.Y.} is grounded in four canonical data sources maintained independently of the institutional network fabric (Fig.~\ref{fig:bodypipeline}, leftmost column). The GLPI~\cite{glpi} asset registry, enriched with MAC
addresses and IP assignments exported from the DHCP lease database, provides the authoritative device inventory via \texttt{profiles\_cam/}. The sovereign DHCP resolver (\texttt{DNSMasq}) binds each registered MAC to a fixed IP
address and a positional hostname encoding campus, block, and floor. Switch hardware profiles (\texttt{profiles\_sw/}) consolidate model, firmware, serial number, PoE budget, and LLDP neighbor data collected during onboarding. Finally, \texttt{topo\_map.yml} encodes the campus-to-switch hierarchy, allowing the entire graph to be rebuilt deterministically from a single script invocation without querying the network. \textsc{B.O.D.Y.} addresses this gap not by replacing existing tools but by fusing what they already know into a deterministic topology graph. 

\subsection{Collection \& Normalization}
\label{sec:collection}

For each managed block-level switch, \texttt{switch\_collect.py} establishes a remote session adapted to the target vendor and extracts the current hardware state: MAC forwarding table, interface status vector, PoE
telemetry, and LLDP neighbor advertisements. The collected output is parsed into normalized structures with a  vendor-agnostic schema and persisted as \texttt{\_profile.json}, decoupling the collection phase from all downstream processing and ensuring that the source of truth remains accessible even during switch unavailability. \textsc{B.O.D.Y.} delivers two complementary topology views from this data: a server and infrastructure view consolidating all campus blocks, and a switch and camera view resolving physical endpoint distribution across floors and ports within each block.

\subsection{Port Classification}
\label{sec:port_class}

To produce a topologically useful representation, knowing which MAC addresses are present on a switch is necessary but insufficient: the operator needs to know, for each port, which device is connected, what its function is, and
where it sits in the physical hierarchy. In ordinary enterprise environments this problem is intractable without SNMP cooperation, because a MAC address alone carries no semantic content. \textsc{B.O.D.Y.} resolves it through corroborating evidence: each independent signal narrows the hypothesis space, and their convergence produces a classification whose confidence grows from probabilistic to deterministic.

The classification begins with LLDP neighbor identity, the strongest available signal and the primary mechanism for anchoring a switch within the spine-leaf hierarchy. Where LLDP neighbor names typically follow a consistent
naming convention, their structure alone discriminates uplinks toward the distribution layer from managed cascades toward the access layer, without requiring any prior knowledge of the physical wiring. This naming convention is deployment-specific and does not require a fixed format: any systematic distinction between upstream and downstream names is sufficient. Where LLDP is absent or suppressed, MAC density per uplink port serves as the structural fallback: the distribution switch, by definition, aggregates the MAC tables of all downstream access switches, so the port carrying the largest MAC population is the uplink, regardless of vendor or protocol support.

With the switch's spine-leaf position established, endpoint classification proceeds through further corroboration. PoE delivery state and MAC density discriminate the remaining port types: a gigabit port negotiating PoE with multiple MACs is an unmanaged mini-switch cascade; a single-MAC port delivering power is a camera endpoint. Each hypothesis is then cross-referenced against \texttt{DNSMasq} conf files and Zabbix query device name: the positional hostname convention encodes campus, block, and floor directly in the device name, so a MAC that resolves to a registered entry confirms not only device identity but physical location. OUI prefix lookup within the constrained manufacturer subspace corroborates device class independently of the registry, and the wattage signature of a PoE port provides a further independent confirmation against known camera model consumption profiles. When the sovereign DHCP registry confirms the MAC, the classification becomes deterministic. Ports
that exhaust all signals without convergence are not silently discarded but exposed as unresolved nodes, preserving full observability of registry integrity and triggering the HIL protocol described in Section~\ref{sec:hil}.

\subsection{Graph Building \& Resolution}
\label{sec:graph_build}

A classified port list transforms into an operational reasoning tool only when organized as a directed graph that propagates failures upward, enables lateral comparison between peers, and directs physical intervention with
spatial precision. The per-switch tree structure (\texttt{\_tree.json}) encodes this directed graph as a declarative filesystem hierarchy where parent-child relationships are expressed through directory nesting~\cite{morris2020, hellerstein2010}. This design choice eliminates external database dependencies, ensures the source of truth remains accessible during upstream failures, and allows the entire graph to be rebuilt deterministically from a single script invocation without querying the network. With the graph structure in place, the next challenge is spatial aggregation: distributing hundreds of endpoints across a flat port list tells the operator nothing about which floor a fault is on.

The declarative \texttt{\_tree.json} hierarchy is machine-consumable by design: a downstream reasoning agent such as B.R.A.I.N.~\cite{brain2026}  can ingest port-to-device bindings, floor groupings, PoE state, and upstream dependencies without network access, enabling physical co-dependency reconstruction and fault propagation tracing grounded  in immutable physical reality rather than probabilistic approximation.

\subsubsection{Floor Grouping and OUI Resolution}
\label{sec:flopr_group}

Spatial precision is achieved by grouping endpoints into logical floor nodes through \texttt{extract\_floor()}, which parses the floor index directly from the positional hostname convention without requiring a manual floor registry. OUI prefix lookup within the constrained manufacturer subspace of the monitored infrastructure provides an independent corroboration of device class, making the assignment deterministic even for endpoints whose hostnames have not yet been registered. These logical floor blocks, illustrated in Fig.~\ref{fig:topologybraincams}, also serve as the substrate for health propagation: verdicts are projected onto the graph through a bottom-up coloring scheme in which a degraded camera propagates a warning signal upward through its floor group to the parent switch, collapsing what would otherwise be dozens of independent alerts into a single spatially-grounded signal. Not all endpoints, however, resolve cleanly through this cascade, and those that do not must be handled without degrading the integrity of the graph.

\begin{figure*}
	\centering
	\includegraphics[width=1\linewidth]{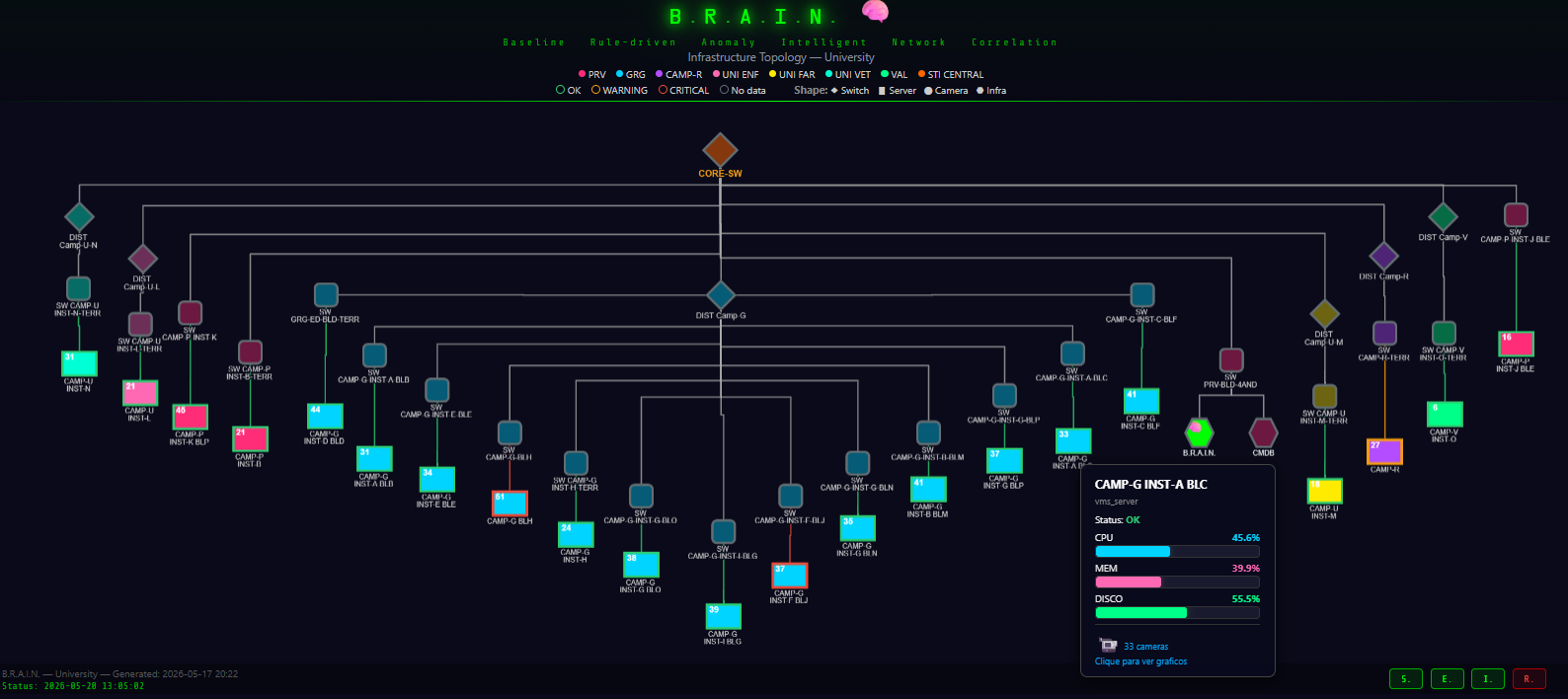}
	\caption{B.R.A.I.N. server and infrastructure topology 		($\mathcal{G}_{srv}$) for a multi-campus video monitoring infrastructure. 	The graph is rooted at the core aggregation switch and expands downward 		through campus-level distribution switches and block-level access switches 		to monitored VMS server nodes. Node fill color encodes campus affiliation; 		border color encodes the health verdict derived from the monitoring platform (green: OK, amber: WARNING, red: CRITICAL). Each server node displays its
	active camera stream count, providing immediate visibility into load distribution across campuses. On hover, real-time CPU, memory, and disk utilization are displayed inline; clicking navigates to the full metric 		history in the monitoring platform. S.E.I.R. Layout persistence controls (bottom right) allow saving, exporting, importing, and resetting the graph arrangement to accommodate different display environments.}
	\label{fig:topologybrainservers}
\end{figure*}

\subsubsection{Asset Integrity Loop}
\label{sec:asset_loop}

Endpoints that exhaust the classification cascade without a registry match are not suppressed but exposed as named nodes carrying port number, OUI prefix, and parent switch identity, equipping the field technician with
maximum available context before any physical intervention is required. This explicit exposure is the visible face of a deeper enforcement mechanism: the sovereign DHCP resolver denies IP addresses to unregistered
MACs, lease durations progress from 12 to 48 hours before stabilization to discriminate transient from permanent endpoints, and periodic MAC surveillance triggers structured alerts through the Zabbix~\cite{zabbix}
monitoring platform and the \textsc{B.R.A.I.N.} cognitive layer~\cite{brain2026}. Together, these mechanisms ensure that the graph remains a faithful representation of the physical infrastructure rather than a static snapshot that drifts from reality over time.

\begin{figure*}[ht]
	\centering
	\includegraphics[width=1\textwidth]{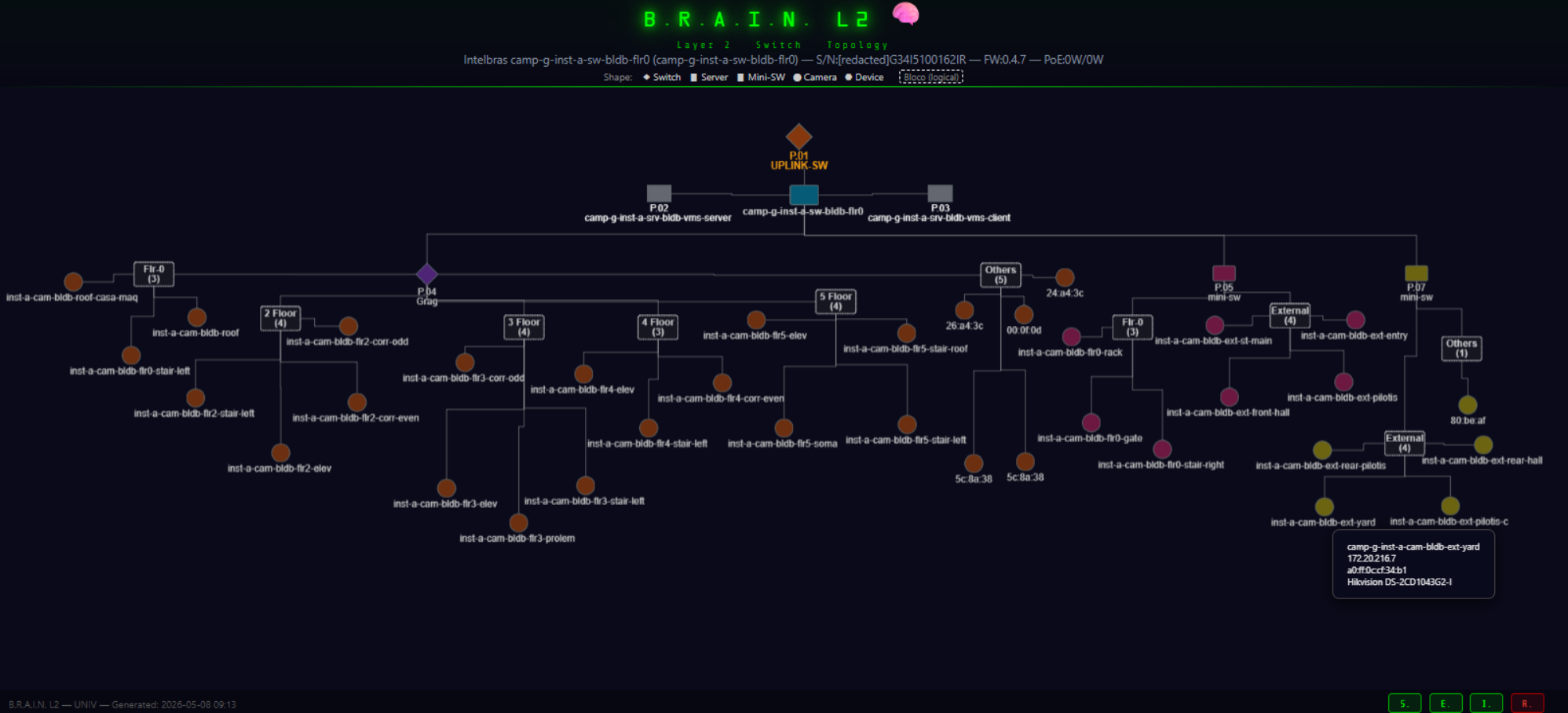}
	\caption{Port-level switch and camera topology for \texttt{camp-g-inst-a-sw-bldb-flr0} 		(Intelbras, 35 cameras). Cameras are grouped into logical floor blocks extracted 		from the positional hostname convention; one uplink port identifies the 		institutional fabric above. A mini-switch cascade (purple diamond) extends 		coverage beyond the managed boundary without participating in any management
	protocol. Some unresolved MACs are exposed in logical topological quarantine block "others"~ carrying OUI prefix 		and parent switch identity. The tooltip overlay (bottom right) illustrates the per-device metadata surfaced on hover: anonymized canonical name, redacted IP address, redacted MAC address, and vendor model string. Note that the floor block is a spatial label derived from naming semantics, not a switching boundary: in larger buildings the same floor may appear under two distinct 	blocks if served by separate managed switches, each defining an independent fault propagation domain. Finally The S/E/I/R controls 	(bottom right) allow saving, exporting, importing, and resetting the layout, supporting adaptation to different screen sizes and publication formats.}
	\label{fig:topologybraincams}
\end{figure*}

The health state of the physical infrastructure is projected onto both graphs through a deterministic coloring function $\mathcal{C}: \mathcal{V} \rightarrow \{\text{Green}, \text{Amber}, \text{Red}\}$, sourced exclusively
from the Zabbix~\cite{zabbix} monitoring platform. Both graphs are structurally static: topology changes only when a new switch is on boarded or an asset is re-registered. What evolves is the health projection. In
$\mathcal{G}_{srv}$ (Fig.~\ref{fig:topologybrainservers}), node fill color encodes campus affiliation while border color encodes the Zabbix health verdict, allowing the operator to distinguish a campus-wide degradation
from an isolated host failure at a glance. Each server node additionally displays its active camera process count, surfacing load imbalances that would otherwise require querying the monitoring platform directly.

In $\mathcal{G}_{sw}$ (Fig.~\ref{fig:topologybraincams}), each device inherits the health color of its parent switch rather than its logical floor block. The floor block is a spatial aggregation derived from the positional hostname convention and groups endpoints by physical location, but connectivity, and therefore fault propagation, follows the switching path. In large buildings, the same floor may be served by more than one managed switch, producing two distinct logical floor blocks at the same level: cameras on the same floor but connected through different switches belong to different propagation domains. This distinction, invisible in a flat alert list, is immediately apparent in $\mathcal{G}_{sw}$: a fault that affects only one floor block while leaving its neighbor healthy points directly to the upstream switch rather than to a building-wide failure. The resulting visual state is auditable and reproducible because the coloring derives exclusively from deterministic Zabbix verdicts: the
same input always produces the same colored graph. Both topology views expose layout persistence controls that allow the operator to save, export, import, and reset the graph arrangement, accommodating the spatial constraints of different display environments without affecting the underlying data. Also unresolved MACs, exposed as "OTHERS" (see Figure \ref{fig:topologybraincams}), a quarantine logical block that aggregates nodes carrying only the OUI prefix and parent switch identity, mark endpoints that require a field visit for physical identification; their presence in $\mathcal{G}_{sw}$ converts an implicit gap in the inventory into an explicit, actionable work order for the field team.

\subsection{Output \& HIL}
\label{sec:hil}

A single topology view covering all assets simultaneously would produce a graph  of several hundred nodes whose cognitive load is disproportionate to any single  investigative task. The operator works in two distinct modes: a systemic mode  that assesses resource health across all campus blocks, and a localized mode  that traces a specific fault to its physical port and floor. B.O.D.Y. partitions  the output accordingly, sizing each view to the operator's working memory during  the corresponding investigative posture.

The pipeline produces two complementary topology views rendered as interactive \textit{Cytoscape.js} graphs, as shown in the Output \& HIL column of Fig.~\ref{fig:bodypipeline}. The switch and camera view
resolves the physical layer at port granularity for each block-level switch, grouping endpoints into logical floor nodes and exposing unresolved MACs explicitly as named nodes. The server and infrastructure view renders the campus-to-switch-to-server dependency chain with health  verdicts sourced from the monitoring platform API and \textsc{B.R.A.I.N.} scores updated daily. The two views are complementary rather than redundant: a health degradation on a server node narrows the investigation to a specific host; the operator then navigates to the
switch view to identify which cameras are affected downstream and on  which floor, closing the causal chain from resource anomaly to physical asset. The topology layer is informative and contextual: it orients the
operator toward probable fault locations but does not replace the monitoring platform, whose metric collection, threshold evaluation, and alerting remain authoritative.

\subsubsection{Human-in-the-Loop Protocol}

Unresolved MACs surfacing in the topology graph trigger a structured HIL protocol. The field technician is dispatched carrying port number, OUI prefix,  parent switch identity, and floor location already known, ensuring that physical  intervention is always evidence-driven rather than exploratory. The onboarding  phase is itself a  HIL checkpoint: the operator validates uplink identification  manually before committing the switch profile, making human accountability structural rather than residual. This design minimizes unnecessary field visits 
while preserving the operator as the final arbiter of decisions that cannot be  resolved remotely.

\section{Implementation and Results}
\label{sec:results}

The B.O.D.Y. architecture was implemented and evaluated across the multi-campus production infrastructure of the Universidade Federal Fluminense (UFF), Brazil, spanning 5 geographically separate physical campuses and 2 isolated auxiliary administrative units. The monitored overlay network operates on a dedicated VLAN administered by the 
surveillance laboratory independently of the institutional switching fabric, instantiating the fragmented administrative boundary scenario described in Section~\ref{sec:related}. The surveillance infrastructure 
has been in continuous operation since 2021 \cite{allan_2024}, accumulating over four years of monitoring data without a formally structured topological representation. B.O.D.Y. was deployed retroactively over this live production environment: the full onboarding pipeline, from asset registry consolidation to interactive topology generation across all reachable campus segments, was completed in approximately one month without service interruption and without requiring cooperation from the institutional network administration.

\subsection{Resolution Accuracy}

The full registered asset base comprises 541 devices encoded in the sovereign DHCP resolver and the GLPI asset registry. As detailed in Table~\ref{tab:results}, the deterministic pipeline resolved 530 of these assets without requiring field intervention or administrative privileges on the institutional routing backbone, yielding a resolution accuracy of 97.9\%. The 11 unresolved entries were traced entirely to cameras absent from the asset registry due to human oversight during the onboarding process, confirming that the resolution ceiling is bounded by registration discipline rather than by any limitation of the method itself. Notably, resolution operates against the static asset registry rather than against the live network state: assets on campus segments currently unreachable due to VPN routing constraints or pending switch onboarding are correctly resolved from stored profiles, demonstrating that the pipeline is resilient to partial network availability.

\begin{table}[ht]
	\centering
	\caption{Topology Resolution Success Across UFF Campuses}
	\label{tab:results}
	\begin{tabular}{lrrr}
		\toprule
		\textbf{Campus Unit} & \textbf{Registered Assets} & \textbf{Resolved} & \textbf{Accuracy (\%)} \\
		\midrule
		Campus G & 320 & 315 & 98.43\% \\
		Campus P & 110 & 104 & 94.54\% \\
		Campus R & 60  & 60  & 100.00\% \\
		Campus V & 20  & 20  & 100.00\% \\
		Auxiliary Units & 20  & 40  & 100.00\% \\
		\midrule
		\textbf{Total Ecosystem} & \textbf{530} & \textbf{541} & \textbf{97.96\%} \\
		\bottomrule
	\end{tabular}
\end{table}

\subsection{Deployment Coverage}

At the time of writing, more than 540 cameras and 30 servers are actively monitored across the reachable campus segments. The switch onboarding pipeline has been executed across 26 managed switches spanning more tha 5 distinct manufacturers, among others: Intelbras, HP/Aruba, Tenda, and Ubiquiti, validating the multi-vendor normalization layer. Each onboarding cycle, including automated CLI collection and HIL-assisted uplink validation, completes in approximately 30 minutes per switch. 

\subsection{Port Classification Results}


Table~\ref{tab:cameras} summarizes the resolution outcome across the 541 unique MAC addresses visible in camera, cascade, and downlink ports of the 26 onboarded switches. The pipeline resolved 530 endpoints (97.96\%), of which 191 are directly powered by the managed switch via PoE and 339 are reachable through unmanaged cascade segments where power is delivered by intermediate mini-switches invisible to the managed switching fabric. This distribution validates the mini-switch dilemma reconstruction, the majority of edge camera deployments rely on unregistered PoE hubs inserted by field technicians to extend coverage, and B.O.D.Y. reconstructs these segments deterministically without requiring the intermediate switches to participate in any management protocol.

\begin{table}[ht]
	\centering
	\normalsize
	\caption{Camera Endpoint Resolution Summary}
	\label{tab:cameras}
	\begin{tabular}{lrr}
		\toprule
		\textbf{Category} & \textbf{Count} & \textbf{(\%)} \\
		\midrule
		Resolved, direct PoE         & 191 & 36.0\% \\
		Resolved, not PoE            & 339 & 63.9\% \\
		\midrule
		\textbf{Total resolved}      & \textbf{530} & \textbf{95.1\%} \\
		\midrule
		Unregistered devices (HIL)   & 11  & 0.02\%  \\
		Unknown devices (HIL)        & 0  & 0\%  \\
		\midrule
		\textbf{Total visible MACs}  & \textbf{530} & \textbf{100\%} \\
		\bottomrule
	\end{tabular}
\end{table}

The remaining 11 MACs (6.2\%) activate the HIL protocol. Field investigation classified these as 3 unregistered camera endpoints requiring GLPI and DHCP registration  and unmanaged HP switches acting as intermediate cascade nodes and NVR classified as \texttt{unknown} exhausted all deterministic classification rules and are exposed explicitly in the topology graph as unresolved nodes for HIL resolution. 

\section{Limitations and Constraints}
\label{sec:limitations}

The production deployment documented in this paper required approximately 60 hours of preparatory work before the first deterministic topology could be generated: interviews with field technicians to recover undocumented
wiring decisions, consolidation of asset information scattered across spreadsheets and institutional memory, bulk renaming of assets to conform to the positional hostname convention, and manual verification of the
resulting topology against physical reality. This convergence process involved multiple iterations of cross-referencing, correction of duplicate MACs and typographical errors in hostnames. In our defense, Table~\ref{tab:cameras} confirms that the number of assets requiring physical field visits remained below 2.5\% of the total visible MAC population. Furthermore, three campus segments remain topologically dark at the time of writing, pending resolution of VPN routing constraints, remote firewall policies, and administrative boundaries with independently operated campuses. These constraints are not limitations of the method itself but of the institutional context in which it was deployed.

The design choices documented in this paper carry additional explicit trade-offs whose boundaries should be understood before the system is adopted or extended.

\subsection{Naming convention dependency.} The positional naming convention enables topology inference without SNMP, network probes, or cooperation from any network element, a capability that no discovery-based approach 
can guarantee in overlay environments with heterogeneous administrative boundaries. The trade-off is that an asset registered with a hostname that deviates from the convention, whether by operator error, legacy naming, or third-party provisioning, falls outside the inferred topology. The system does not fail silently: non-conforming assets are flagged as unresolved nodes in the topology graph. 

\subsection{Managed switch dependency.} The pipeline requires at least one managed, SSH-accessible switch per monitored network segment. Segments served exclusively by unmanaged switches cannot be on boarded, and their 
endpoints remain invisible to the topology graph until a managed switch is introduced. This constraint is an intentional economic design choice: a single managed switch per block is sufficient to resolve the entire 
downstream segment, keeping deployment costs low while preserving full topological coverage of the reachable infrastructure.

\subsection{HIL onboarding dependency.} Switch onboarding requires manual operator intervention to validate uplink identification before the switch profile is committed. This makes human accountability structural rather 
than residual, but introduces an operational dependency: onboarding throughput is bounded by operator availability, and campus segments remain topologically dark until their block-level switch is onboarded. 
In the current deployment, three campus segments remain pending due to VPN routing constraints and physical switch access limitations rather than any limitation of the method itself.

\subsection{Registry passivity.} The pipeline resolves device identity against the sovereign DHCP resolver and the GLPI asset registry, both of which reflect the state of the infrastructure at the time of registration. A surveillance infrastructure in continuous operation since 2021 \cite{allan_2024} accumulates a registration debt: devices added, replaced, or decommissioned outside the formal registration workflow produce MACs that are visible in the switch tables but absent from the registry. In the current deployment, 33 such MACs were identified across the 26 onboarded switches, of which 20 are unregistered camera endpoints, 10 are unmanaged intermediate switches, and 3 are devices of undetermined type. The system exposes these as HIL candidates rather than suppressing them, but resolving the registration debt requires field intervention that the pipeline cannot automate.

\subsection{LLDP coverage.} Port classification relies on LLDP neighbor advertisements to deterministically identify uplinks and managed cascades. Switches that do not implement LLDP, or that implement non-standard variants, fall back to heuristic classification based on MAC density and PoE state. In the current deployment, this fallback produces 16 ports classified as \texttt{unknown}, predominantly servers connected on non-gigabit interfaces without active PoE negotiation. The ongoing deployment of \texttt{lldpd} on monitored Linux hosts is 
expected to eliminate this residual class in subsequent onboarding cycles.

\subsection{Generalizability.} The evaluation results characterize pipeline behaviour within a single deployment unit, a Brazilian federal university surveillance network in continuous operation since 2021. No claim of 
population-level generalisability is made. A deployment on a different fleet, with different vendor compositions and administrative boundary structures, would produce different resolution rates and HIL candidate 
distributions. What the evaluation guarantees is reproducibility: given the same asset registry, the same switch profiles, and the same deterministic classification rules, any operator can re-run the pipeline and obtain identical results.

The positional hostname convention is simultaneously the enabling mechanism and the principal operational dependency of B.O.D.Y.: naming discipline replaces protocol-level cooperation, and the system fails gracefully rather than silently when that discipline lapses. Unresolved assets are exposed as named HIL candidates rather than suppressed, bounding the failure mode to inventory gaps rather than topological hallucinations. The same structural constraint applies to any overlay fleet on a dedicated VLAN without uniform SNMP cooperation: wireless access point networks such as eduroam, industrial IoT fabrics, VoIP endpoint networks, physical access 
control systems, and building automation segments all share the same precondition — a managed switch per segment and a registration discipline that encodes physical location in device identity.

\subsubsection{Generalization to Corporate Environments}

Corporate network environments relax several of the constraints that motivated B.O.D.Y.'s design. Administrators typically hold SNMP read access across the switching fabric, reducing the dependency on non-privileged CLI collection. IT service management platforms such as GLPI, ServiceNow, and SCCM maintain asset registries whose agents periodically report hostnames, MAC addresses, and software inventory, providing an independent source of naming truth that can be cross-referenced against the sovereign DHCP resolver. Endpoint protection platforms similarly maintain per-host identity records that survive naming convention violations. In these environments, B.O.D.Y.'s multi-modal fusion pipeline can be extended to consume these additional signals, reducing 
dependence on strict naming discipline without abandoning the deterministic classification cascade.

\subsubsection{Reducing Naming Dependency via Protocol-Level Identity}

Where administrative control permits, deploying LLDP agents on monitored Linux and Windows hosts via configuration management tools such as Ansible eliminates the naming convention dependency for uplink and cascade classification entirely: LLDP neighbor advertisements carry host identity independently of hostname conventions, and their presence in the switch MAC table provides a deterministic classification signal that survives legacy naming 
debt. Bulk hostname remediation, distributed via Ansible playbooks or group policy, converts legacy naming debt into conformant positional identifiers without requiring physical intervention. Together, these mechanisms demonstrate that although naming convention constitutes a single point of failure in constrained overlay environments, it is a tractable one: with moderate operational discipline and standard configuration management 
tooling, the conditions for deterministic topological grounding are achievable in any network where at least one managed switch per segment is accessible.

\section{Conclusion}
\label{sec:conclusion}

B.O.D.Y. addresses a problem that the topology discovery literature has not previously framed: maintaining deterministic, auditable knowledge of physical network structure in overlay infrastructures where administrative cooperation cannot be assumed and where the monitored asset base accumulates a registration debt over years of 
continuous operation. By fusing MAC forwarding tables collected via direct CLI sessions on sovereign access switches with PoE telemetry, LLDP neighbor advertisements, OUI fingerprinting, and a sovereign DHCP resolver, the pipeline resolves device identity without dependency on institutional SNMP cooperation or upstream administrative access.

The production deployment across five campuses of the Universidade Federal Fluminense, conducted over approximately one month on a surveillance infrastructure in continuous operation since 2021, resolved 524 camera endpoints across 24 onboarded switches with a global asset registry accuracy of 99.03\%. Unresolved MACs are not 
suppressed but exposed as structured HIL candidates, equipping field technicians with port number, MAC addres, and parent switch identity before any physical intervention is required. The declarative filesystem hierarchy ensures that the source of truth remains accessible and auditable regardless of network availability, while 
the bottom-up graph coloring scheme propagates health verdicts from individual endpoints upward through logical floor groups to the campus-level dependency chain.

B.O.D.Y. is the structural foundation of the H.U.M.A.N. framework. The deterministic topology it produces eliminates the topological hallucinations that degrade probabilistic AIOps inference engines when decoupled from physical ground truth, creating the immutable grounding layer on which the B.R.A.I.N. \cite{brain2026} cognitive inference layer operates. The method generalizes to any overlay network operating on a dedicated VLAN with at least one managed switch per segment, encompassing surveillance networks, IoT deployments, VoIP infrastructures, and enterprise wireless networks operating under fragmented administrative boundaries.

The grounding layer established by B.O.D.Y. and the diagnostic compass provided by B.R.A.I.N. together constitute a necessary but incomplete AIOps loop: the operator remains the sole agent capable of acting on the produced hypotheses. The third component of the H.U.M.A.N. framework, H.A.N.D.S., will close this loop by introducing supervised machine access to monitored hosts, enabling evidence-driven remediation actions to be proposed, validated by the operator, and executed directly against the physical infrastructure. This progression from passive grounding to active agency under human supervision defines the long-term trajectory of the framework.

\balance

\bibliographystyle{IEEEtran}
\bibliography{body_refs}

@article{notaro2021survey,
  author    = {Notaro, Paolo and Cardoso, Jorge and Gerndt, Michael},
  title     = {A Survey of {AIOps} Methods for Failure Management},
  journal   = {ACM Trans. Intell. Syst. Technol.},
  volume    = {12},
  number    = {6},
  pages     = {81:1--81:45},
  year      = {2021},
  doi       = {10.1145/3464906}
}

@article{ahmat_topology_survey,
  title={Ethernet topology discovery: A survey},
  author={Ahmat, Kamal},
  journal={arXiv preprint arXiv:0907.3095},
  year={2009}
}

@inproceedings{farkas2008topology,
  author    = {J{\'a}nos Farkas and Vinicius Garcia de Oliveira and
               Marcos Rog{\'e}rio Salvador and Giovanni Curiel dos Santos},
  title     = {Automatic Discovery of Physical Topology in {Ethernet} Networks},
  booktitle = {Proc. IEEE AINA},
  year      = {2008},
  pages     = {848--854}
}

@article{hallucination_survey_2025,
  title={A survey on hallucination in large language models: Principles, taxonomy, challenges, and open questions},
  author={Huang, Lei and Yu, Weijiang and Ma, Weitao and Zhong, Weihong and Feng, Zhangyin and Wang, Haotian and Chen, Qianglong and Peng, Weihua and Feng, Xiaocheng and Qin, Bing and others},
  journal={ACM Transactions on Information Systems},
  volume={43},
  number={2},
  pages={1--55},
  year={2025},
  publisher={ACM New York, NY}
}

@article{gobjuka2010,
  title={Ethernet topology discovery for networks with incomplete information},
  author={Gobjuka, Hassan and Breitbart, Yuri J},
  journal={IEEE/ACM Transactions on Networking},
  volume={18},
  number={4},
  pages={1220--1233},
  year={2010},
  publisher={IEEE}
}

@book{morris2020,
  title={Infrastructure as code: Dynamic systems for the cloud age},
  author={Morris, Kief},
  year={2025},
  publisher={" O'Reilly Media, Inc."}
}

@article{hellerstein2010,
  title={The declarative imperative: experiences and conjectures in distributed logic},
  author={Hellerstein, Joseph M},
  journal={ACM SIGMOD Record},
  volume={39},
  number={1},
  pages={5--19},
  year={2010},
  publisher={ACM New York, NY, USA}
}

@misc{zabbix,
  author = {Zabbix SIA},
  title = {Zabbix Documentation},
  year = {2026},
  url = {https://www.zabbix.com/documentation/},
  note = {Accessed: May, 2026}
}

@misc{glpi,
  author       = {{Teclib'}},
  title        = {{GLPI - Gestionnaire Libre de Parc Informatique}},
  year         = {2026},
  url          = {https://glpi-project.org/},
  note         = {Accessed: 2026-05-03}
}

@inproceedings{rca2024automatic,
  title={Automatic root cause analysis via large language models for cloud incidents},
  author={Chen, Yinfang and Xie, Huaibing and Ma, Minghua and Kang, Yu and Gao, Xin and Shi, Liu and Cao, Yunjie and Gao, Xuedong and Fan, Hao and Wen, Ming and others},
  booktitle={Proceedings of the Nineteenth European Conference on Computer Systems},
  pages={674--688},
  year={2024}
}

@inproceedings{Breitbart2004NetInventory, title = {Topology discovery in heterogeneous IP networks: the NetInventory system}, author = {Breitbart, Y. and Garofalakis, Minos N. and Jai, Ben and Martin, Clifford E. and Rastogi, R. and Silberschatz, A.}, booktitle = {IEEE/ACM Transactions on Networking}, year = {2004}, month = {06}, }

@article{Pandey2011SNMPTopology, title = {SNMP-based enterprise IP network topology discovery}, author = {Pandey, Suman and Choi, Mi-Jung and Won, Young J. and Hong, J. W.}, journal = {International Journal of Network Management}, year = {2011}, month = {05}, }

@article{Sun2006CRTTopology, title = {A Method of Topology Discovery for Switched Ethernet Based on Address Forwarding Tables}, author = {Sun, Yan-tao and Wu, Zhi-mei and Shi, Zhiqiang}, journal = {Journal of Software}, year = {2006}, }

@article{Son2005MetroEthernet, title = {Physical Topology Discovery for Metro Ethernet Networks}, author = {Son, Myunghee and Joo, Bheom-soon and Kim, Byungchul and Lee, Jae-Yong}, journal = {ETRI Journal}, year = {2005}, month = {08}, }

@article{Tao2009LinkLayerSNMP, title = {Link Layer Topology Discovery Algorithm Based on SNMP}, author = {Tao, J.}, journal = {Computer Engineering}, year = {2009}, }

@inproceedings{Wang2016SNMPAlgo, title = {An algorithm and implementation of network topology discovery based on SNMP}, author = {Zhangchao, Wang and Yan, Zhang and De-dong, Zhang and Hongwei, W.}, booktitle = {2016 First IEEE International Conference on Computer Communication and the Internet (ICCCI)}, year = {2016}, month = {10}, }

@article{Zhang2017SNMPOptimization, title = {An Optimization Algorithm of Network Topology Discovery Based on SNMP Protocol}, author = {Zhang, Xuliang}, journal = {Journal of Computational Chemistry}, year = {2017}, month = {12}, }

@misc{niedermaier2019,
  author    = {Niedermaier, Matthias and Hanka, Thomas and Plaga, Sven and von Bodisco, Alexander and Merli, Dominik},
  title     = {Efficient Passive {ICS} Device Discovery and Identification by {MAC} Address Correlation},
  year      = {2019},
  howpublished = {arXiv preprint arXiv:1904.04271},
  url       = {https://arxiv.org/abs/1904.04271}
}

@inproceedings{hadzic2001,
  author    = {Hadzic, Ilija},
  title     = {Hierarchical {MAC} Address Space in Public {Ethernet} Networks},
  booktitle = {Proc. IEEE Global Telecommunications Conference (GLOBECOM'01)},
  volume    = {3},
  pages     = {1563--1569},
  year      = {2001},
  month     = {November},
  organization = {IEEE},
  note      = {Cat. No. 01CH37270}
}

@inproceedings{imamura2020,
  title={A device identification method based on combination of multiple information},
  author={Imamura, Yuki and Nakamura, Nobuyuki and Yao, Taketsugu and Ata, Shingo and Oka, Ikuo},
  booktitle={NOMS 2020-2020 IEEE/IFIP network operations and management symposium},
  pages={1--4},
  year={2020},
  organization={IEEE}
}

@unpublished{brain2026,
  author    = {de Sousa, Cledson},
  title     = {From Metrics to Root Cause: {B.R.A.I.N. — A}
Baseline Rule-driven Anomaly Intelligent Network
Correlation},
  note      = {First installment of the \textsc{H.U.M.A.N.} framework
               trilogy. Manuscript in preparation},
  year      = {2026}
}

@article{lowekamp2001topology,
  title={Topology discovery for large ethernet networks},
  author={Lowekamp, Bruce and O'Hallaron, David and Gross, Thomas},
  journal={ACM SIGCOMM Computer Communication Review},
  volume={31},
  number={4},
  pages={237--248},
  year={2001},
  publisher={ACM New York, NY, USA}
}

@INPROCEEDINGS{allan_2024,
  author={dos Santos, Allan Costa Nascimento and de Paula, Karina and Vidal, Marcos T. L. and da Silva, João M. M. and de Sousa, Cledson and Fernandes, Leandro A. F. and de Castro, Tiago B. and Bedo, Marcos and Kohwalter, Troy C. and Bastos, Carlos A. M. and Seixas, Flavio L. and Fernandes, Natalia C. and Muchaluat-Saade, Débora C. and Ghinea, Gheorghita},
  booktitle={2024 IEEE International Conference on E-health Networking, Application and Services (HealthCom)}, 
  title={A Computer Vision Model to Support Individuals with Disabilities Within University Campuses}, 
  year={2024},
  volume={},
  number={},
  pages={1-7},
  doi={10.1109/HealthCom60970.2024.10880838}
}

@article{hong2025llm_netmgmt,
  author    = {Hong, Jin-Keun},
  title     = {A Comprehensive Survey on {LLM}-Based Network Management
               and Operations},
  journal   = {International Journal of Network Management},
  volume    = {35},
  number    = {6},
  pages     = {e70029},
  year      = {2025}
}

@inproceedings{donadel2024llms,
  author    = {Donadel, Denis and Marchiori, Francesco and Pajola, Luca and Conti, Mauro},
  title     = {Can {LLMs} Understand Computer Networks? {Towards} a Virtual System Administrator},
  booktitle = {Proc. IEEE 49th Conference on Local Computer Networks (LCN)},
  year      = {2024},
  doi       = {10.1109/LCN60385.2024.10639641}
}

@inproceedings{genet2025,
  author    = {Ifland, Beni and Krief, Rubin and Zilberman, Aviram and Duani, Elad
               and Ohana, Miro and Murillo, Andres and Manor, Ofir and Lavi, Ortal
               and Hikichi, Kenji and Shabtai, Asaf and Elovici, Yuval and Puzis, Rami},
  title     = {{GeNet}: A Multimodal {LLM}-Based Co-Pilot for Network Topology and Configuration},
  booktitle = {Proc. IEEE 45th Int. Conf. Distributed Computing Systems Workshops (ICDCSW)},
  year      = {2025},
  doi       = {10.1109/ICDCSW63273.2025.00026}
}

@inproceedings{hamadanian2023holistic,
  author    = {Hamadanian, Pouya and Arzani, Behnaz and Fouladi, Sadjad
               and Kakarla, Siva Kesava Reddy and Fonseca, Rodrigo
               and Billor, Denizcan and Cheema, Ahmad and Nkposong, Edet
               and Chandra, Ranveer},
  title     = {A Holistic View of {AI}-Driven Network Incident Management},
  booktitle = {Proc. 22nd ACM Workshop on Hot Topics in Networks (HotNets)},
  pages     = {180--188},
  year      = {2023},
  doi       = {10.1145/3626111.3628176}
}
\end{document}